\documentclass[aps, prd, twocolumn, nofootinbib,longbibliography]{revtex4-1}

\usepackage{epsfig}
\usepackage{slashed, enumitem}
\usepackage[utf8]{inputenc}

\usepackage{float,color}
\usepackage{subfigure}
\usepackage{multirow}
\usepackage{slashed}
\usepackage{placeins} 
\usepackage{mathrsfs} 

\usepackage{amsmath}
\usepackage{amssymb}
\usepackage{graphicx}

\usepackage{slashed}
\usepackage{multirow}
\usepackage[dvipsnames]{xcolor}
\usepackage{epstopdf}
\usepackage{soul}
\usepackage{tikz}

\usepackage{siunitx}
\usepackage{xspace}

\usepackage{subfigure}

\usepackage{upgreek}
\usepackage{mathtools}
\usepackage{setspace, slashed, gensymb, hhline, booktabs}
\usepackage{wrapfig}
\usepackage{makecell}

\usepackage{mhchem}

 \usepackage[colorlinks=true, citecolor=blue,linkcolor=blue]{hyperref}

\newcommand{\el}[2]{$^{#1}\text{#2}$}

\usepackage{tikz,xcolor,hyperref}

\definecolor{lime}{HTML}{A6CE39}
\DeclareRobustCommand{\orcidicon}{\hspace{-4pt}
	\begin{tikzpicture}
		\draw[lime, fill=lime] (0,0) 
		circle [radius=0.16] 
		node[white] {\hspace{0.1mm}{\fontfamily{qag}\selectfont \tiny ID}};
		\draw[white, fill=white] (-0.07,0.1) 
		circle [radius=0.01];
	\end{tikzpicture}
	\hspace{-3.2mm}
}

\foreach \x in {A, ..., Z}{\expandafter\xdef\csname 
	orcid\x\endcsname{\noexpand\href{https://orcid.org/\csname orcidauthor\x\endcsname}
		{\noexpand\orcidicon}}
}

\begin{document}
	
\onecolumngrid
\begin{flushright}
TIFR/TH/21-20
\end{flushright}
\twocolumngrid

%\title{ {\sc Large Energy Singles} from Atmospheric Neutrinos and Dark Matter}

 \title{Large Energy Singles at JUNO from Atmospheric Neutrinos and Dark Matter}

\author{Bhavesh Chauhan\orcidA{} }
\author{Basudeb Dasgupta\orcidB{}}
\author{Amol Dighe\orcidC{}}
\affiliation{Tata Institute of Fundamental Research, Homi Bhabha Road, Mumbai 400005, India}

\date{\today}

\begin{abstract}
Large liquid scintillator detectors, such as JUNO, present a new opportunity to study 
neutral current events from the low-energy end of the atmospheric neutrinos, and possible 
new physics signals due to light dark matter. We carefully study the possibility of detecting  
``Large Energy Singles'' (LES), i.e., events with visible scintillation energy $>15$\,MeV, but no 
other associated tags. For an effective exposure of 20 kton-yr and 
considering only Standard Model physics, we expect the LES sample 
to contain $\sim40$ events from scattering on free protons and $\sim 108$ events from interaction 
with carbon, from neutral-current interactions of atmospheric neutrinos. Backgrounds, largely due to 
$\beta$-decays of cosmogenic isotopes, are shown to be significant only below 15 MeV visible energy. 
The LES sample at JUNO can competitively probe a variety of new physics scenarios, such as 
boosted dark matter and annihilation of galactic dark matter to sterile neutrinos. 
\end{abstract}

\maketitle

\section{Introduction}
Atmospheric neutrinos are produced in the interactions of cosmic rays with Earth's atmosphere. 
Measurements of these atmospheric neutrinos at detectors such as  Super-Kamiokande 
have been crucial for the discovery of neutrino oscillations\,\cite{Super-Kamiokande:1998kpq}. 
Despite extraordinary achievements over several decades, the detection of low-energy non-electron 
neutrino flavours, i.e., $\nu_\mu$, $\bar{\nu}_\mu$, $\nu_\tau$, and $\bar{\nu}_{\tau}$, has remained 
elusive essentially because at Cherenkov detector like Super-Kamiokande, such a detection depends 
on having a charged particle above the Cherenkov threshold\,\cite{Beacom:2003zu}.

Scintillator detectors do not require charged particles to cross the Cherenkov threshold for them to be 
detected. In particular, neutral-current interactions such  as $\nu + p \rightarrow \nu + p$ lead to a 
prompt visible scintillation, which can be detected even for neutrinos with energies of tens of 
MeV\,\cite{Beacom:2002hs}. 
The difficulty is that the signal has a single component, as opposed to inverse beta decays where a 
neutron tag is possible in addition to the initial prompt scintillation from the charged lepton. The 
``singles" from neutrino sources can be mimicked by other processes, and therefore the backgrounds 
are usually quite large. Indeed, for small 
detectors, the singles from atmospheric neutrinos are often considered as a  
background\,\cite{Atroshchenko:2016bpy, 
KamLAND:2011fld}. 
However, upcoming large volume liquid scintillator detectors, such as JUNO (Jiangmen Underground 
Neutrino 
Observatory)\,\cite{JUNO:2015zny}, will accumulate a significant number of such singles, which may allow a first measurement of the low-energy end of the $\nu_\mu$, $\bar{\nu}_\mu$, $\nu_\tau$, and $\bar{\nu}_{\tau}$ atmospheric neutrino spectra.

%However, at large detectors like JUNO these 
%events can contribute a large sample over the course of a few years. 

One may ask what sources and interaction channels could lead to such singles. In a 
scintillator detector, the events below a few MeV visible energy will be dominated by intrinsic 
radioactivity. Between 5-15 MeV, the events are dominated by decays of cosmogenic 
isotopes. At ``large" visible energies, i.e., above 15\,MeV, the events are dominated by neutral-current 
interactions of atmospheric neutrinos. We propose that JUNO maintain a {Large 
Energy Singles ({\sc LES})} database comprising of singles with visible energy $E_{\rm vis}\in (15, 
\,100)\,{\rm MeV}$, 
which will contain evidence of neutral-current interactions of atmospheric neutrinos, and possibly even 
of interesting physics beyond the standard model. 

In this paper, we predict the {\sc LES} spectrum at JUNO. We identify the main contributions to the signal in \S\,\ref{sec:signal}, study the dominant backgrounds at low energy and estimate the threshold from a veto analysis in \S\,\ref{sec:bkg}, and present our main result in \S\,\ref{sec:res}. Further in \S\,\ref{bdm}, we explore well-motivated new physics scenarios that can have a visible imprint in the JUNO {\sc LES} data. For example, we discuss the sensitivity to boosted dark matter and annihilation of galactic dark matter to sterile neutrinos. We end the 
paper with a brief summary and outlook in~\S\,\ref{conc}.

%In section\,\ref{sec:signal}, we estimate the energy distribution of 
%singles at JUNO from the contributing channels. In section\,\ref{sec:bkg}, we 
%study the cosmogenic backgrounds and determine the 
%threshold for the LES sample. The main result of this paper, the LES spectrum at JUNO, is 
%presented in section\,\ref{sec:res}. We 
%provide the projected sensitivity to some new physics scenarios in section\,\ref{bdm}. We end the 
%paper with a brief summary and outlook in section\,\ref{conc}.

\section{Singles at JUNO}\label{sec:signal}

The LES events from atmospheric neutrinos arise mostly from elastic scattering with protons ($\nu p$ 
ES), and 
quasi-elastic-like scattering with carbon ($\nu \rm C$ QEL) which results in single or multiple proton 
knockouts.  
The scintillation signal from elastic scattering and ``proton-only" knockouts cannot be distinguished, 
and the 
detector only measures the sum of these two channels.

The neutral-current interactions of neutrinos are sensitive to all flavors; however, they do 
not distinguish between flavors. Therefore, the only measurable quantity is the spectrum of the sum of 
events from all flavors. In general, the differential event rate with respect to the recoil energy of the 
proton ($T_p$) is given by
\begin{equation}
	\frac{dN}{dT_p} =  N_t~T~\sum_{f}\int dE_\nu \frac{d\phi^f}{dE_\nu} \frac{d\sigma^f}{dT_p}\,,
\end{equation}
where $f\in \{\nu_e, \bar{\nu}_e\, \nu_\mu, \bar{\nu}_\mu, \nu_\tau, \bar{\nu}_\tau \}$, $N_t$ is the 
number of targets, and $T$ is the data-taking time 
period which we have considered to be 1 yr, unless specified. For a 20\,kton fiducial volume detector, 
the number of target protons from hydrogen (i.e., free protons) is $N_p 
=1.5\times10^{33}$ and the number of target carbon nuclei is  $N_C 
=8.8\times10^{32}$\,\cite{JUNO:2015zny}.  

\subsection{Atmospheric neutrino fluxes}

%The atmospheric neutrino spectrum  in the range 
%$ E_\nu\in (100\,{\rm MeV},\,10\,{\rm TeV})$, measured by Super-Kamiokande\,\cite{Super-Kamiokande:2015qek}, is 
%consistent with the detailed simulations performed by Honda et al.\,\cite{Honda:2015fha} 
%and the neutrino oscillation hypothesis. The detected events arise primarily from charged current 
%interactions. However, the flux of $\nu_\mu$, $\bar{\nu}_\mu$, $\nu_\tau$, and $\bar{\nu}_{\tau}$ 
%below 200\,MeV can only be detected via neutral current, and LES at JUNO is a promising prospect. 

%Super-Kamiokande has measured the atmospheric neutrino spectrum in the energy range 
%100\,MeV - 10\,TeV\,\cite{Super-Kamiokande:2015qek} which is consistent with the detailed 
%simulations performed by Honda et al.\,\cite{Honda:2015fha} and the neutrino oscillation hypothesis. 
%As a large fraction of events arise from charged current interactions, the lower energy threshold is 
%determined by the detectability of Cherenkov radiation from recoiling lepton i.e., 100\,MeV for 
%electron-flavor, 200\,MeV 
%for muon-flavor, and 3.5 GeV for tau-flavor neutrinos and antineutrinos. The flux of $\nu_\mu$, 
%$\bar{\nu}_\mu$, $\nu_\tau$, and $\bar{\nu}_{\tau}$ 
%below 200\,MeV can only be detected via neutral current, and LES at JUNO is a promising prospect. 

The flux of atmospheric neutrinos for $E_\nu>100$\,MeV at the site of JUNO is calculated in 
Ref.\,\cite{JUNO:2021tll} based on  the predictions of Honda et al.~\cite{Honda:2015fha}. As it is 
located at a lower latitude than Super-Kamiokande, it is estimated that the atmospheric neutrino 
flux at JUNO is $\sim10\%$ smaller\,\cite{JUNO:2021tll}. 

The atmospheric neutrino flux for $E_\nu<100$\,MeV has been 
determined by the FLUKA group for Super-Kamiokande and Borexino\,\cite{Battistoni:2005pd}. 
There are large uncertainties ($\sim25\%$) in the flux prediction originating from 
the dependence on the geomagnetic field. The 
predicted flux for $E_\nu<100$\,MeV at JUNO can be approximately obtained by scaling the 
Super-Kamiokande prediction by a factor of 0.9. 

For our analysis, we use the scaled Honda et al. fluxes 
above 100 MeV and scaled FLUKA fluxes below 100 MeV with appropriate matching. Our simplified 
estimates agree with the predictions in Ref.\,\cite{JUNO:2021tll}.

\subsection{Singles from $ \nu p$\,ES}

\subsubsection{$\nu p$\,ES cross section}

The $\nu p$ ES cross section is a robust prediction of the Standard Model and has been measured 
experimentally\,\cite{Ahrens:1986xe}. The differential cross section for this process in 
terms of the neutrino energy ($E_\nu$) and recoiling proton kinetic energy ($T_p$) is given as
\begin{equation}\label{diffxs}
\frac{d\sigma}{dT_p} = \frac{G_F^2 M_p^3}{4 \pi E_\nu^2} \left[	A \pm B \frac{s-u}{M_p^2} + C \frac{(s-u)^2}{M_p^4}\right]\,,
\end{equation}
where $M_p$ is the mass of proton, $s-u = 4 M_p E_\nu - 2 M_p T_p$, while the functions 
$A,~B,~\rm{and}~C$ depend on 
$E_\nu$, the 
momentum transfer ($Q^2 = 2 M_p T_p$), and form factors of proton. The expressions 
can be 
found in Ref.\,\cite{Ahrens:1986xe}. A more familiar expression for low-energy interactions, in the 
small-$Q^2$ limit, is given in Ref.\,\cite{Beacom:2002hs}. We retain the $Q^2$-dependence in our 
analysis, but note that the impact on the event rates is small. 

Due to the large energy loss rate of proton, the scintillation from a recoiled proton is nearly 
isotropic and the direction of proton (and hence, 
that of the incident neutrino) is not reconstructed. As a result, the angular distribution of these events 
is not measurable and 
we, therefore, only focus on the angle-averaged cross section and flux.

The main uncertainty in the differential cross section originates from the 
uncertainties associated with the axial-mass parameter $M_A$ in the axial form-factor. There are 
contributions from strange sea-quark to all of the form factors, however these have not been taken 
into account for 
our estimations. For this analysis, we fix $M_A = 1.03$\,GeV \cite{Bernard:2001rs}, and conservatively 
assume that the cross section 
uncertainties are 
$\mathcal{O}(10\%)$.

For neutrinos and anti-neutrinos with $E_\nu\geq550\,\rm MeV$, the momentum transfer to 
proton is large enough for the single-pion production through the delta resonance. However, we 
expect 
such processes to have much smaller cross sections than $\nu p$ ES, and the corresponding 
event rates can be ignored. The $\nu e$ ES cross section is relatively smaller, and we can safely 
ignore these interactions in our analysis.

\subsubsection{Quenched proton scintillation}

Due to photosaturation losses, i.e., 
quenching, the visible energy ($E_{\rm vis}$) is different from the kinetic energy of the recoiling 
proton. The differential event spectrum in terms of visible energy is given by 
\begin{equation}\label{eq:q}
	\frac{dN}{dE_{\rm vis}} = \left(\frac{dE_{\rm vis}}{dT}\right)^{-1}\frac{dN}{dT}.
\end{equation}
The visible energy is related to $T_p$ through
\begin{equation}
	E_{\rm vis}(T_p) = \int_{0}^{T_p} \frac{dT}{1 + k_B \langle dE/dx \rangle + k_C \langle dE/dx 
		\rangle^2 }\,,
\end{equation}
where $k_B = 6.5\times10^{-3}$ g/cm$^2$/MeV and $k_C = 1.5\times 10^{-6} 
$ (g/cm$^2$/MeV)$^2$ are Birks' constants\,\cite{Birks:1951boa}. The average energy loss 
during propagation, $\langle dE/dx \rangle$, is determined by the baseline 
parameters\footnote{$\langle 
	dE/dx \rangle$ $\approx$ $0.88\,\langle dE/dx\rangle_C + 0.12\, 
	\langle dE/dx\rangle_H$.}\textsuperscript{,}\footnote{The energy loss rate on carbon and hydrogen 
	can be obtained 
	from 
	\url{www.physics.nist.gov/PhysRefData/Star/Text/PSTAR.html}.} of JUNO simulations\,
\cite{JUNO:2015zny}. Note that, the non-linear but one-to-one mapping between $E_{\rm vis}$ and 
$T_p$, and good energy resolution of the detector imply that the effects of quenching can be inverted 
and one can, in principle, obtain 
$dN/dT_p$ from $dN/dE_{\rm vis}$. This is useful in reconstructing the incident neutrino spectrum 
and has been studied in the context of supernova neutrinos\,\cite{Dasgupta:2011wg,  Li:2019qxi}.
 
 \subsubsection{Predicted $\nu p$\,ES spectrum}
  
\begin{figure}[t]
	\centering
	\includegraphics[width=\columnwidth]{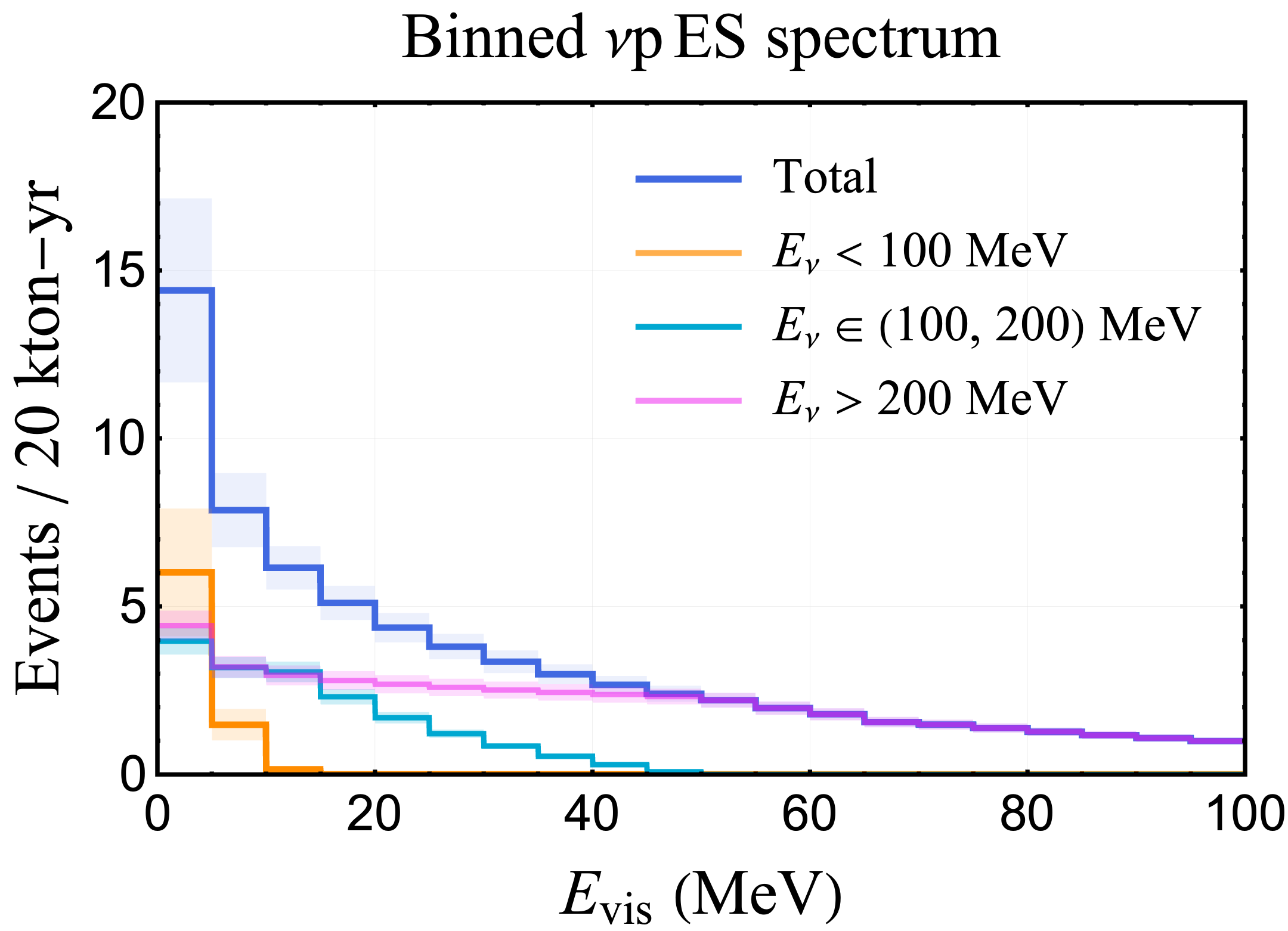}
	\caption{\label{fig:flux_split} The binned event spectrum ($\Delta E_{\rm bin} = 5$ MeV) from 
	three energy ranges of the atmospheric neutrino spectrum, and the total spectrum. The 
	shaded 
	regions represent the uncertainties arising from the cross section and flux estimates. Note that 
	the uncertainties are larger in the lower energy bins, mostly because of the uncertain neutrino 
	flux for $E_\nu<$\,100\,MeV.}
\end{figure}

The atmospheric neutrinos with $E_\nu \in(100,\,200)\,\rm MeV$ are special, because only the 
electron-flavor component has 
been measured through charged current interactions. We ask if the muon and tau flavor components 
have a detectable imprint on $\nu p$ ES events spectrum. For this purpose, we take a closer look at 
the distribution of events from various parts of 
the atmospheric neutrino spectrum, and divide the flux into three energy ranges:
\begin{enumerate}
	\item $E_\nu<100$ MeV, which has large uncertainties,
	\item $E_{\nu} \in$\,(100, 200)\,MeV which is partly measured, and
	\item $E_\nu > 200$ MeV, which is well determined.
\end{enumerate}
The event spectrum from these three energy ranges and the total event spectrum, is shown in 
Fig.\,\ref{fig:flux_split}. We also show the uncertainty in the event spectrum from cross section as well 
as flux estimates.

The events from an incident neutrino with 
energy $E_\nu$ are 
distributed over the range $ 0 \leq 
T_p\leq 2 M_p E_\nu/(M_p + E_\nu)^2$. As a result, there is a significant contribution to low-energy 
bins from the high energy part of the atmospheric neutrino spectrum. It seems, a priori, that 
reconstruction of the incident spectrum from $\nu p$ ES events will be challenging.

From Fig.\,\ref{fig:flux_split}, one notes that in the visible energy range $E_{\rm 
vis}$\,$\in$\,(15,\,40)\,MeV, the 
contribution from neutrinos with $E_\nu \geq 200 \, \rm MeV$ is similar for all the energy bins, whereas 
the contribution from $E_{\nu}$\,$\in$\,(100,\,200)\,MeV decreases with energy. With 
much larger exposure, this excess of events at low energies can become statistically significant, and if 
detected, would represent a 
measurement of the flux of $\nu_\mu + \bar{\nu}_\mu + \nu_\tau + \bar{\nu}_\tau$ for $E_\nu \leq200 
\, \rm MeV$, after 
statistically 
subtracting the contribution from $\nu_e + \bar{\nu}_e$. One can also note that the contribution 
from the flux with $E_\nu < 100 \, \rm MeV$, which has large uncertainties, is mostly in the low energy 
bins, and will not be relevant if we have an energy threshold of $E_{\rm vis}\sim15$\,MeV.

\subsection{Singles from $\nu \rm C$\,QEL}

%$\approx (3\,\rm fm)^{-1}$
 
The neutral-current interactions of atmospheric neutrinos in JUNO have been studied in 
 Ref.\,\cite{Cheng:2020aaw}. These interactions are dominated by quasi-elastic-like (QEL) processes 
 where one or more nucleons can be knocked out of \el{12}{C}. A detailed study of this process has 
 been 
 carried out in Ref.\,\cite{Cheng:2020aaw}, which reports the event rates for various channels ($1p,~ 
 1n, ~1p1n, ~2p,~ 2n,$ ...), as well as the recoil proton spectrum from the sum of these channels. The 
 total event rate with at least one proton in the final state\,\footnote{This can be obtained by adding 
 histograms given in 
 Fig.\,4 of 
 Ref.\,\cite{Cheng:2020aaw} as well as integrating the spectrum in Fig.\,6 of 
 Ref.\,\cite{Cheng:2020aaw}.} is 
 found to be $\sim 30\,\rm kton^{-1}yr^{-1}$. For 20 kton-yr exposure of JUNO, this implies an 
 aggregate of $\sim$600 events, distributed over $T_p \in $(0.1, 300)\,MeV. 
 
We need to estimate the singles event rate from $\nu \rm C$ QEL process, which arises from the single 
proton knockout
\begin{equation}
 1p :\qquad 	\nu + \ce{^{12}_{}C} \rightarrow \nu + p  + \ce{^{11}_{}B}^{(\ast)}\,,\quad
\end{equation}
and from multiple proton knockouts such as
\begin{eqnarray}
	2p:\qquad \nu + \ce{^{12}_{}C} &\rightarrow \nu + 2p + \ce{^{10}_{}Be}^{(\ast)}\,, \\
	3p:\qquad \nu + \ce{^{12}_{}C} &\rightarrow \nu + 3p + \ce{^{9}_{}Li}^{(\ast)}.
\end{eqnarray}
These protons are a part of the total proton spectrum given in Ref.\,\cite{Cheng:2020aaw}. To isolate 
the singles events, we calculate the fraction of ``proton-only" knockout events that do not have a 
neutron in the final state. Using the results in  
 Ref.\,\cite{Cheng:2020aaw}, we find that
 \begin{equation}
 	\frac{N_{1p} + N_{2p} + N_{3p} + ... }{ N_{1p} + N_{1p1n} + N_{2p} + N_{1p2n} + N_{2p1n} + ... } 
 	\approx 
 	0.52,
 \end{equation}
which implies that roughly half of the protons do not have a neutron tag. Therefore, the singles 
spectrum from $\nu \rm C$ QEL interaction can be approximated by scaling the proton spectrum 
given in \cite{Cheng:2020aaw} by 0.52. The $E_{\rm vis}$ distribution is obtained by applying the 
effects of quenching using Eq.\eqref{eq:q}. 
 
The daughter nuclei in a QEL process can de-excite through protons and/or alpha particle emission. 
However, these particles are 
estimated to be lower in number, and most of them have kinetic energies below 20 MeV 
\cite{Cheng:2020aaw}. 
Considering the effects of quenching, these particles will be below threshold, and hence can be 
ignored. Moreover, after de-excitation, some of the channels result in unstable nuclei with short 
lifetimes that undergo $\beta$ decay. These $\beta$ decays can be used to tag the proton 
scintillation 
events, as has been demonstrated in Ref.\,\cite{Cheng:2020oko}. While this will allow us to identify a 
fraction of the LES sample as coming from $\nu \rm C$\,QEL, we do not use this information in our 
analysis. It is also possible that the elastic scatterings between a knockout neutron and free 
proton in detector results in visible scintillation. However, these events will be vetoed by the 
accompanying neutron capture.
 
\begin{figure}[t]
	\centering
	\includegraphics[width=\columnwidth]{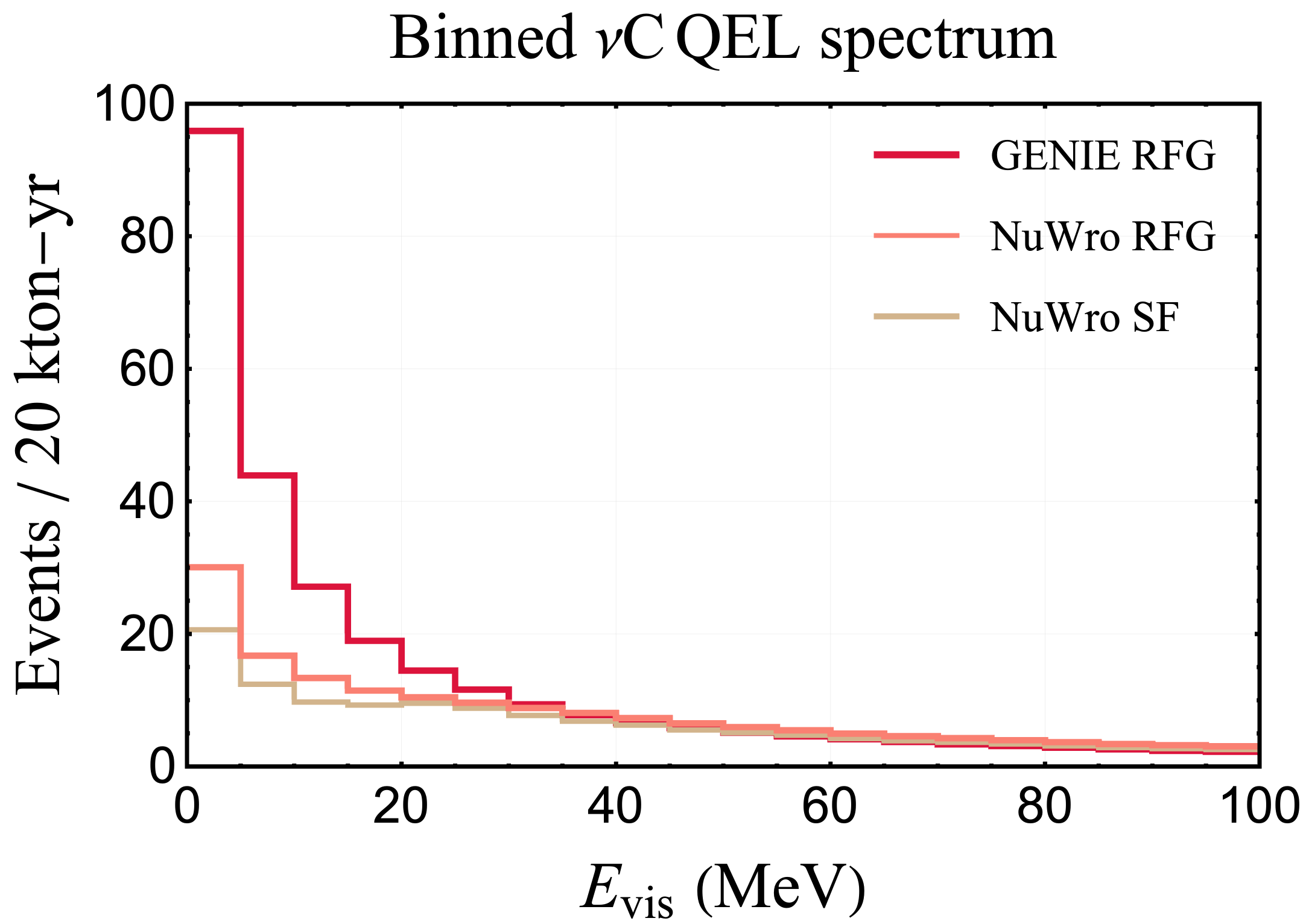}
	\caption{\label{fig:qel} The binned energy distribution of singles ($\Delta E_{\rm bin} = 5$ 
	MeV) from $\nu \rm C $ QEL interactions is shown for different neutrino 
		interaction MC generators (GENIE, and NuWRO with nuclear structure models -- RFG and SF). }
\end{figure}

The event rate predictions in Ref.\,\cite{Cheng:2020aaw} depend on the choice of Monte Carlo 
generator for neutrino interactions. In Fig.\,\ref{fig:qel}, we show the expected rate of $\nu \rm C $ QEL 
singles at JUNO, as predicted by the neutrino event generators GENIE~\cite{Andreopoulos:2009rq} 
and NuWro~\cite{Golan:2012rfa}. It appears that $\nu \rm C$ QEL predictions are sensitive to details of 
the nuclear structure, unlike the robust predictions for $\nu p$ ES. For our analysis in the rest of the 
paper, we use the results obtained by GENIE~\cite{Andreopoulos:2009rq}, with 10\% uncertainty. 

\section{Backgrounds and threshold }\label{sec:bkg}

\subsection{Cosmic muon spallation}
 
The passage of cosmic muons through the detector produces isotopes through spallation. These 
unstable isotopes decay in the detector, and their daughter particles can lead to visible signals. The 
singles background (i.e., without an associated neutron capture) originates 
from $\beta^\pm$, $\beta^\pm \gamma$, $\beta^\pm p$, and $\beta^\pm \alpha$ decays of these 
cosmogenic isotopes. The 
$\beta n$ decays do not  contribute to singles as the neutron can be tagged. 

The cosmic muon spallation and isotope yields have been extensively studied 
for Super-Kamiokande in Refs.\,\cite{Li:2014sea, Li:2015kpa, Li:2015lxa}. The liquid scintillator detector 
KamLAND has measured the yields of some cosmogenic isotopes \cite{KamLAND:2009zwo}.  Since 
the average 
muon energy at the JUNO site is lower than that at KamLAND\footnote{$\langle E_\mu\rangle\sim$ 215 
GeV for JUNO 
	\cite{JUNO:2015zny}, 
	and $\langle E_\mu\rangle\sim$ 260 GeV for KamLAND\,\cite{KamLAND:2009zwo}.}, the isotope 
	yields 
at JUNO would be nearly 90\% of that at KamLAND \cite{JUNO:2015zny}. In this paper, we 
scale the measured KamLAND yields to JUNO where available, and for other cosmogenic isotopes, we 
use simulation yields from Table 13-9 in Ref.\,\cite{JUNO:2015zny}. The 
isotope yields and other details are given in Table\,\ref{tab}.

\subsubsection{Decay of cosmogenic isotopes}

To predict the visible-energy distribution of singles from cosmogenic isotopes, we estimate their 
production rate in the detector. Looking at their half-lives (cf., Table\,\ref{tab}), it is reasonable to 
assume that all the cosmogenic isotopes would decay within a day. The production count per day 
(CPD) of the radio-isotope $i$ is given as
\begin{equation}\label{eq:cpd}
\text{CPD}_i = Y_i R_\mu T \rho \langle L_\mu \rangle\,,
\end{equation}
where $Y_i$ is the isotope yield, $R_\mu$ = 3\,Hz is the rate of cosmic ray muons traversing through 
in JUNO, $T$ = 86400\,s is 
the time interval, and $\langle L_\mu \rangle \approx $  23\,m is the average muon track length in 
JUNO\,\cite{JUNO:2015zny}. These values give a more useful and simplified expression
\begin{equation}
\text{CPD}_i = \frac{47.7}{\rm day} \left( \frac{Y_i}{10^{-7}{\rm muon}^{-1}{\rm g}^{-1}\text{cm}^2}\right)\,,
\end{equation}
which we use in the evaluation of the event rate $R_i = \mathcal{B}_i\times\rm{CPD}_i$, 
as well as the event spectrum
\begin{equation}
\frac{dN_i}{dE_{\rm{vis}}} = \mathcal{B}_i\times{\rm{CPD}_i} \times{f_i(E_{\rm{vis}})}\,.
\end{equation}
Here $\mathcal{B}_i$ is 
the branching ratio of the isotope to the singles channels, and $f_i$ is the normalized 
 distribution of $E_{\rm vis}$ for the $i^{\rm th}$ isotope\,\footnote{The normalized distribution can be 
 obtained from 
\url{www-nds.iaea.org/relnsd/vcharthtml/VChartHTML.html} which uses BetaShape 
\cite{Mougeot:2015bva}.}. As the individual isotopes cannot 
be identified, only the cumulative event spectrum from these isotopes can be measured.

\subsubsection{Veto criterion}
The cosmogenic isotopes that decay within a few seconds of the muon passage can be 
tagged, and the events can be removed by imposing appropriate spatial and temporal cuts. This was 
demonstrated for 
Super-Kamiokande in Ref.\,\cite{Li:2015kpa}. By accounting for the muon energy 
deposition along the track, it was proposed to veto a small cylindrical volume centered around the 
muon track. A similarly detailed analysis for JUNO is required, but is beyond the scope of this paper. 
To get rid of cosmogenic isotope decays, we propose a much more conservative veto --- 
a cylindrical volume around the entire track of the cosmic muon with radius $R_{\rm 
veto}$ for a time $\Delta t_{\rm veto}$. This results in a dead volume fraction of
\begin{equation}
	\frac{\delta V}{V} \sim \frac{(R_\mu\,\Delta t_{\rm veto})\times \pi R_{\rm veto}^2 
	\langle L_\mu \rangle }{\frac43 
	\pi R_{\rm CD}^3 }\,,
\end{equation}
where $R_{\rm CD} = 17.7$\,m is radius of the central detector and $\langle L_\mu \rangle \approx 
$  23\,m is the average muon track length in 
JUNO\,\cite{JUNO:2015zny}. For $R_{\rm veto} = 3$ m and 
$\Delta t_{\rm veto} = 1.2$ s (similar to Ref.\,\cite{JUNO:2015zny}), the dead volume fraction is 
$\sim$10\%, whereas for $\Delta t_{\rm veto} = 2$ s, 
the dead volume fraction increases to $\sim$17\%. We envision that this dead volume fraction of 
detector will be compensated by longer duration of data taking, to get the appropriate effective 
exposure.  

%We provide our estimates for an effective exposure of 20\,kton-yr.

Note that, these estimates do not account for the muon tagging and track reconstruction 
efficiencies. A detailed Monte Carlo simulation, which also accounts for detector systematics, is 
required to optimize the veto criterion. In Ref.\,\cite{JUNO:2015zny}, similar cuts were proposed to 
identify the inverse beta-decay events in JUNO, with additional cuts to tag the neutron capture.

\begin{figure}[t]
	\centering
	\includegraphics[width=\columnwidth]{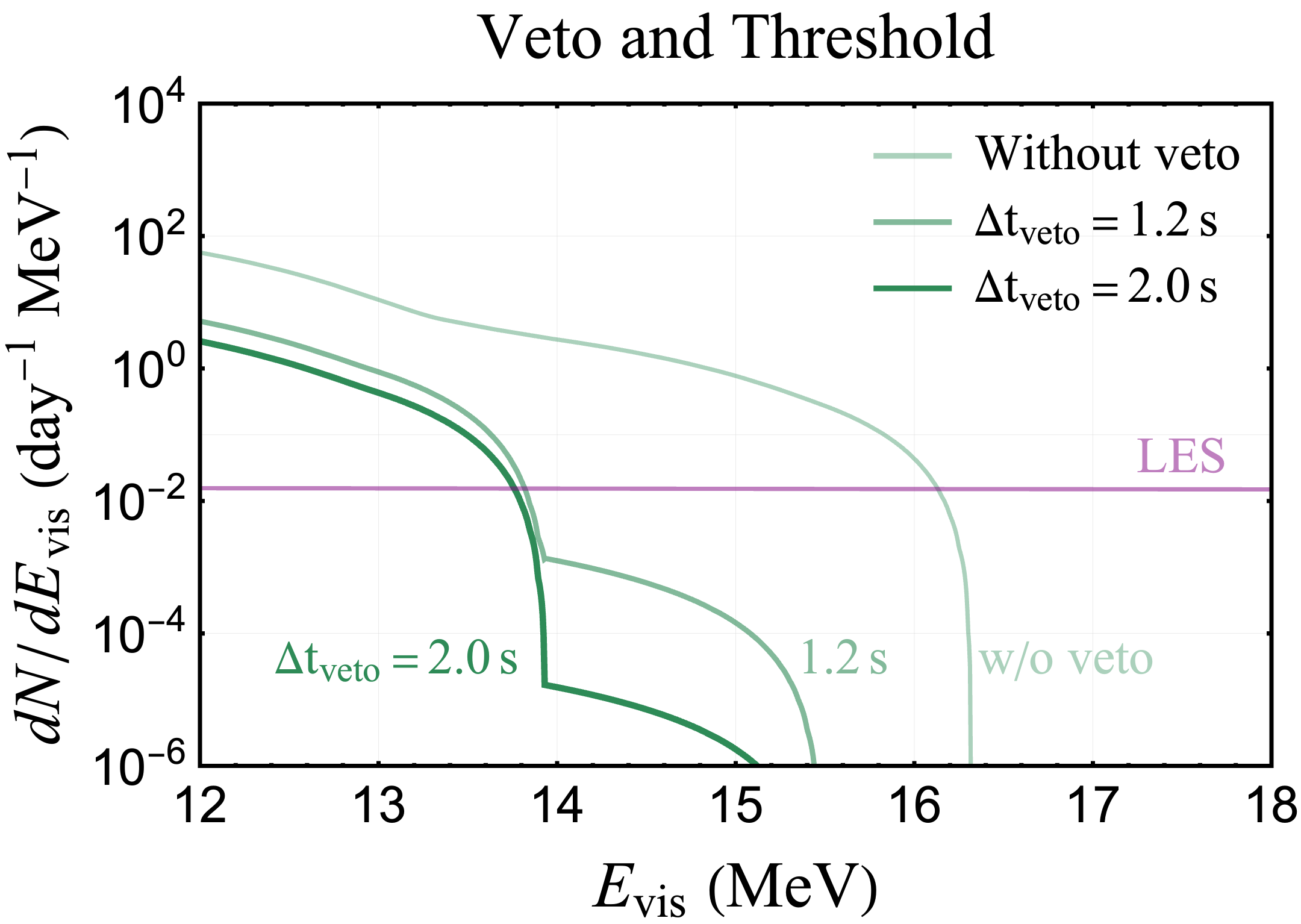}
	\caption{\label{fig:threshold}  The cumulative event spectrum from the cosmogenic isotope 
		decay for $\Delta t_{\rm veto}$ = 0\,s (equivalent to without veto), 1.2\,s, and 2.0\,s. We  show 
		the LES spectrum with light-purple curve for comparison. We will take $\Delta t_{\rm veto}$ = 
		2.0\,s as our benchmark.}
\end{figure}

\subsubsection{Irreducible background and threshold}
The fraction of cosmogenic isotopes that decay outside the $\Delta t_{\rm veto}$ window cannot 
be tagged, and constitute the irreducible background. The 
effective rate of background events from a cosmogenic isotope is given by
\begin{equation}\label{eq:rtil}
 \tilde{R}_i =  e^{- \Delta t_{\rm veto} / \tau_i} \times \mathcal{B}_i\times 
\text{CPD}_i\,,
\end{equation}
which is provided in Table\,\ref{tab}. The effective event spectrum is estimated by
\begin{equation}
	\frac{dN_i}{dE_{\rm{vis}}} = e^{- \Delta t_{\rm veto} / \tau_i} \times \mathcal{B}_i\times{\rm{CPD}_i} 
	\times{f_i(E_{\rm{vis}})}.
\end{equation}
In Fig.\,\ref{fig:threshold}, we have shown the cumulative event spectrum from 
cosmogenic isotope decay for $\Delta t_{\rm veto}$ = 0\,s (equivalent to without veto), 1.2\,s, and 
2.0\,s. Without any veto, the cosmogenic isotope decays constitute a \emph{wall}-like background 
at $E_{\rm vis}\sim$16.5 MeV. Our veto criterion, with $\Delta t_{\rm veto}$ = 2\,s, allows lowering the 
threshold up to 14 MeV. We present our results with a conservative threshold of 15 MeV. Note that, 
by considering the energy deposition along the track, the length of the 
cylindrical volume veto can be reduced and $\Delta t_{\rm veto}$ can be increased with little change to 
dead volume fraction. This significantly reduces the cosmogenic backgrounds\,\cite{Li:2015kpa}.

\begingroup
\renewcommand{\arraystretch}{1.5}
\begin{table}[t!]
	\centering
	\scriptsize
	\caption{\label{tab} The details of the cosmogenic isotopes considered in this paper is tabulated. 
		The endpoint of the beta spectrum ($E_\beta^{\text{max}}$) and the half-life $T_{1/2}$ were 
		obtained from 
		\url{www-nds.iaea.org}. 
		The experimentally measured yields by KamLAND\,\cite{KamLAND:2009zwo} have been scaled to 
		obtain the yields in JUNO. Wherever KamLAND measurements are not available, results of JUNO 
		simulations\,\cite{JUNO:2015zny} have been used. The isotope production count per day 
		(CPD) from Eq.\,\eqref{eq:cpd} captures the number of isotopes that are produced in 
		the detector per day. The fraction of these isotopes that decay outside the $\Delta t_{\rm veto}$ = 
		2\,s 
		constitute the irreducible background, whose rate $\tilde{R}$ (using Eq.\,\eqref{eq:rtil}) is 
		tabulated.
	} 
	\begin{tabular}{c c c  c  c c c }
		\hline \hline 
         \multirow{3}{*}{\makecell[cc]{Radio\\[4pt]Isotope}}&  
         \multirow{3}{*}{\makecell[cc]{$E_\beta^{\text{max}}$\\[4pt](MeV)}} &  
         \multirow{3}{*}{\makecell[cc]{$T_{1/2}$\\[4pt](s)}}  & 
         \multicolumn{2}{c}{Yield ($10^{-7} 
         \mu^{-1} 
			g^{-1}\text{cm}^2$)}  & \multirow{3}{*}{\makecell[cc]{CPD\\[4pt](per day)}}   & 
			\multirow{3}{*}{\makecell[cc]{$\tilde{R}$\\[4pt](per day)}}  \\  
		 & & 
		&\begin{tabular}{@{}c@{}}for KamLAND \\ 
		(Ref.\,{\cite{KamLAND:2009zwo}})\end{tabular} & \begin{tabular}{@{}c@{}}for JUNO \\ (this 
		work)\end{tabular} & &  \\ 
		\hline
		\el{14}{B} & 20.64 & 0.0126  & -- & $4.4 \times 10^{-3}$ & 0.021 & $\sim0$  \\
		
		\el{12}{N} &16.32 & 0.0011 & $1.8 \pm 0.4$ & $1.62$ & 77.3 & $\sim0$  \\
		
		\el{9}{C} &15.47 & 0.126& $3.0 \pm 1.2$ & $ 2.7 $ & 128.8 & $0.002$ \\
		
		\el{8}{B} & 13.9 & 0.770& $ 8.4 \pm 2.4$ & $ 7.56 $ & 360.6 & $59.61$ \\
		
		\el{9}{Li} & 13.60 & 0.178& $ 2.2 \pm 0.2$ & $ 1.98 $ & 94.4 & $0.019$ \\
		
		\el{13}{B} &13.43 & 0.0174 & -- & 0.251 & 12 & $\sim0$ \\
		
		\el{12}{B} &13.37 & 0.0202 & $  42.9   \pm  3.3  $ & 38.6 & 1841.4 & $\sim0$  \\
		
		\el{8}{Li} &12.97 & 0.839 & $  12.2   \pm  2.6  $ & 10.98 & 523.7 & 100.38 \\
		
		\el{18}{N} &11.92 & 0.62  & -- & $1.88 \times 10^{-4}$ & 0.009 & $0.001$ \\
		
		\el{12}{Be} &11.71 & 0.0215& -- & $9.43 \times 10^{-3}$ & 0.45 & $\sim0$ \\
		
		\el{11}{Be} & 11.51 & 13.76& $  1.1   \pm  0.2  $ & 0.99 & 47.2 & 42.67  \\
		
		\el{16}{N} &10.42 & 7.13 & -- & $0.273$ & 13 & 10.70 \\
		
		\el{15}{C} & 9.77 & 2.449& -- & $1.26 \times 10^{-2}$ & 0.6 & 0.34 \\
		
		\el{8}{He} &9.67 & 0.119 & $  0.7   \pm  0.4  $ & 0.63 & 30.0 & $\sim0$  \\
		
		\el{17}{N} & 8.68 & 4.173& -- & $8.8 \times 10^{-3}$ & 0.42 &0.015 \\
		
		\el{6}{He} &3.51 & 0.80 & -- & 11.40  & 544 & 97.657 \\
		
		\el{10}{C} & 1.91 & 0.747 & $16.5 \pm 1.9$ & 14.85 & 708.35 & 110.77 \\
		
		\el{13}{N} & 1.198 & 597& -- & $0.398$ & 19 &18.96 \\
		
		\el{11}{C} & 0.96 & 1221& $866 \pm 153$ & $ 779 $ & 37177 & 377134 \\
		\hline
	\end{tabular}
\end{table}
\endgroup

\subsection{Other backgrounds}

The other known singles backgrounds include intrinsic radioactivity and reactor neutrinos. 
However, these neutrinos contribute for $E_{\rm vis}\leq$10 MeV. As cosmogenic backgrounds already 
overwhelm the signal at these energies, we do not discuss them in detail. 

Incomplete reconstruction of events, e.g., missing one or more final state particles, can lead to LES 
events. For example, a low-energy $\bar\nu_e$ interacting via charged 
current produces a positron and a neutron; if the neutron is not tagged this can contribute to 
an LES event. However,  for the $E_{\rm vis}\in(15\,-100)\,{\rm MeV}$ window considered for LES events, the relevant 
low-energy $\bar\nu_e$ have a small event rate. Therefore we expect such backgrounds 
to be small. However, a detailed study of the detector efficiencies and reconstruction is warranted.

\section{Forecast for JUNO-LES}\label{sec:res}

\begin{figure}[t!]
	\centering
	\includegraphics[width=\columnwidth]{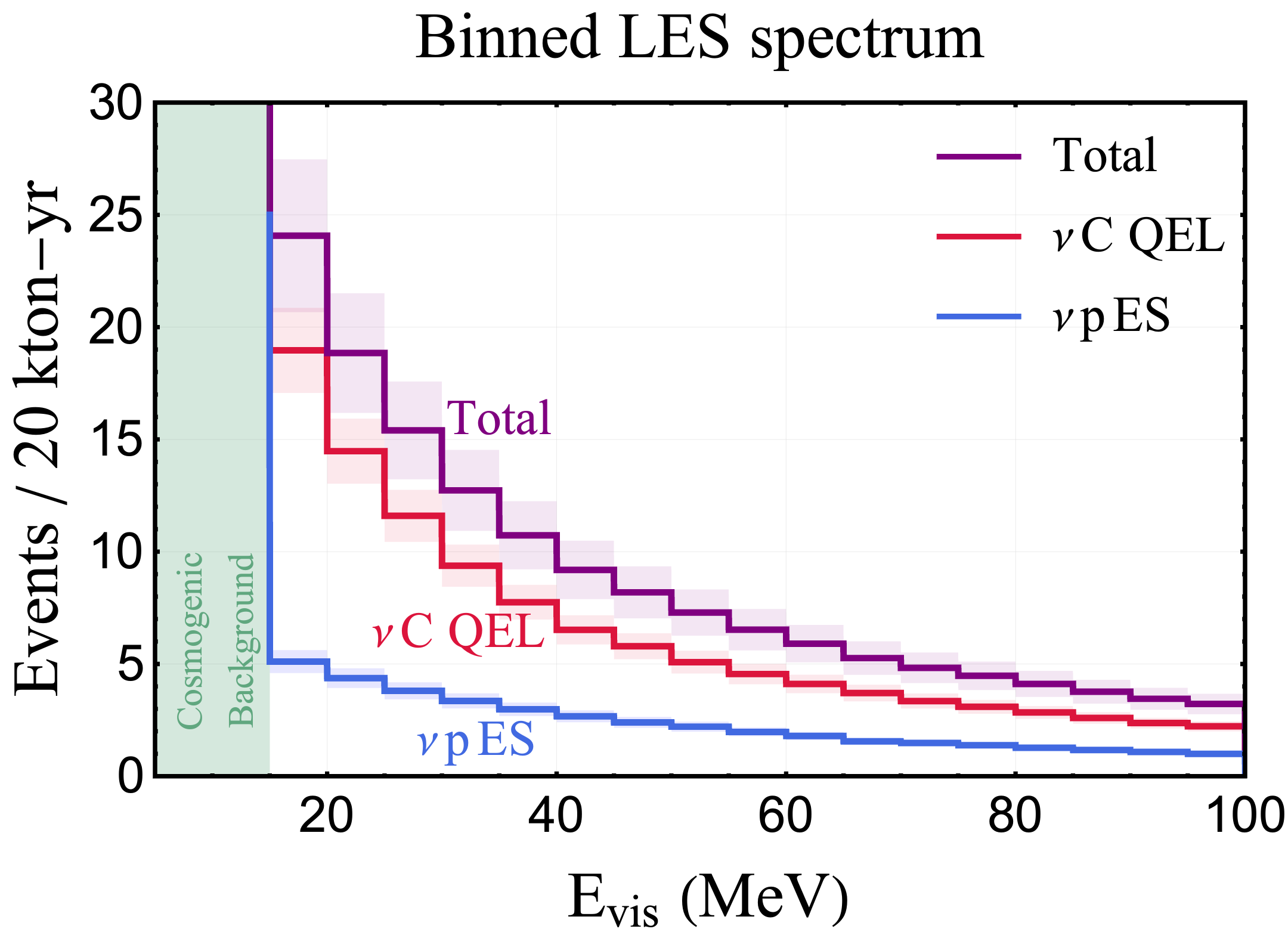}
	\caption{\label{fig:result} The binned event-rate spectrum for $\nu p$ ES (blue) and proton knockouts 
		from $\nu$C QEL interactions (red) are shown. The cumulative spectrum from both the channels 
		is shown in purple. The green shaded region below 15 MeV represents the overwhelming 
		background from cosmogenic isotope decay.}
\end{figure}

It is clear that the low-energy singles spectrum at JUNO will be dominated by irreducible 
cosmogenic isotope decay, intrinsic radioactivity, solar and reactor neutrinos. Above 15 MeV visible 
energy, the events dominantly arise from $\nu p$ ES and $\nu {\rm C}$ QEL 
interactions. This JUNO-LES sample will provide evidence of neutral-current interactions of 
atmospheric neutrinos. In Fig.\,\ref{fig:result}, we have shown 
our estimate for the binned event spectrum for $E_{\rm vis}\in$ (15, 100) MeV from $\nu p$ ES, $\nu 
{\rm C}$ QEL, and the ``Total" sum of the two.  There are a few events expected above 100 MeV, but 
we do not include them in our counting. For 20\,kton-yr exposure, we expect $\sim$40 events from 
$\nu p$\,ES, and $\sim$108 events from $\nu {\rm C}$ QEL. Therefore, we expect a total of $\sim$148 
events with $E_{\rm vis}\in$ (15, 100) MeV in the \mbox{JUNO-LES}\,sample. 

The first goal of JUNO would be to establish the existence of LES events, and therefore, the neutral 
current interactions of atmospheric neutrinos. In this analysis, the backgrounds for $E_{\rm vis} \geq 
15$\,MeV have been assumed to be negligible. Therefore, the first LES events may be observed with a 
few tenths of kton-yr exposure, according to our estimations.  If we further want to claim 
a discovery of $\nu p$ ES events above the ``background" of $\nu \rm C$ QEL events, we need a 
larger exposure. Assuming only statistical errors\footnote{We use the figure of merit $S/\sqrt{B}$ as a 
measure of discovery sensitivity, and $S/\sqrt{S + B}$ to obtain exclusion limits 
\cite{ParticleDataGroup:2020ssz}.} and no other 
background, we estimate that JUNO can discover 
$\nu p $ ES at $3\, \sigma$ ($5\, \sigma$) with 12 (34) kton-yr exposure. Note that, these estimates 
are only based on the counting of events, and more detailed analysis can be performed which also 
accounts for the energy distributions of the $\nu p $\,ES and $\nu \rm C$ QEL events. 

\section{Sensitivity to New Physics}\label{bdm}

Measurement of the {\sc LES} sample at JUNO will open the window to testing many new 
physics scenarios. In this section, we first show that interesting limits can be obtained using 
model-independent analysis. Later, through two examples, we will also show how one can obtain 
model-dependent limits on possible fluxes of energetic new particles.\footnote{Novel interactions can 
also modify 
the 
$\nu p$ ES cross section itself, which we do not study here.}

\subsection{Model independent limits}

A flux of ``Beyond Standard Model'' particles $\psi$ can arise from the annihilation/decay of galactic or 
solar dark matter 
\cite{Murase:2016nwx, McKeen:2018pbb, Kim:2020ipj, Kelly:2019wow}. They can also be emitted 
during evaporation of the primordial black holes\,\cite{Calabrese:2021src, Chao:2021orr}. These 
particles can also be produced in astrophysical processes or through cosmic ray interactions 
\cite{Agashe:2014yua, Bringmann:2018cvk, Cappiello:2018hsu, Ema:2018bih, Alvey:2019zaa, 
Dent:2019krz,Berger:2019ttc, Ge:2020yuf, Cho:2020mnc, Xia:2020apm, Bell:2021xff, Wang:2021nbf, 
Xia:2021vbz, Jho:2021rmn, Das:2021lcr, Ghosh:2021vkt,Wang:2021jic}.  We 
choose to remain independent of the production mechanism and assume a flux of boosted ($E>m$) 
particles with monochromatic energy spectrum. The flux at a detector can be written as
\begin{equation}
	\frac{d\phi_\psi}{dE} = \phi_0\,\delta(E - E_\psi)\,,
\end{equation}
where $\phi_0$ is the normalization in units of $\rm cm^{-2}sec^{-1}$ and $E_\psi$ is the energy of 
the boosted particle. If this flux arises from dark 
matter annihilation ($\rm DM + 
DM \rightarrow \psi + \psi$) or decay ($DM \rightarrow \psi \psi$), then $E_\psi = M_{\rm DM}$ or 
$E_\psi = M_{\rm DM}/2$, respectively.

The boosted particle would be detected through elastic scattering with protons in the detector ($\psi + 
p 
\rightarrow \psi + p$). The cross section for this process depends on the details of the particle physics 
model. We consider two benchmark scenarios for the mediator -- heavy and light, and 
two models for the interaction -- vector and axial vector. 

For the heavy mediator case, one is only 
sensitive to a ratio of coupling strength ($g^{\,}_{\rm H}$) and the mediator mass ($M_{Z^\prime}$). 
On the other hand, for the light mediator case, one is only sensitive to the coupling strength. In order to 
compare 
quantities with the same dimensions, 
for the light mediator scenario we take the heavy scale (corresponding to $ M_{Z^\prime}$ above) to 
be the mass of 
proton. As a result, the strength of interaction in these two scenarios is captured by an effective 
parameter
\begin{equation}\label{eq:geff}
	G_{ \rm eff} = 
	\begin{cases} 
		\begin{aligned}
			\frac{g_{\rm H}^{2}}{8M_{Z^\prime}^2}c_{V/A} &\quad\rm  \,\left[ \, Heavy~mediator\,(HM) \, 
			\right]   
			\\[1ex]
			\frac{g_{\rm L}^{ 2}}{8M_{p}^2}c_{V/A} &\quad\,\rm \left[ \, Light~mediator\,(LM) \, 
			\right]\,,   
		\end{aligned}
		
	\end{cases}
\end{equation}
where $c_{V(A)} = 0.04\,(0.64)$ arises from the form-factors of proton. The differential cross section 
for 
$\psi p$ ES is
\begin{equation}
	\frac{d\sigma_{\psi p}}{dT_p} = 
	\begin{cases}
	\begin{aligned}
		&G_{\rm eff}^2 \frac{M_p}{\pi}\left(1 \pm \frac{M_p 
			T_p}{2E_\psi^2}\right) \hspace{1cm}\quad\rm ~\left[ \, HM \, \right]\\[1ex]
		 &G_{\rm eff}^2 \frac{M_p}{\pi}\left(1 \pm \frac{M_p 
			T_p}{2E_\psi^2}\right)\times\frac{M_p^2}{4 T_p^2}\quad\,\rm \left[ \, LM  \, \right]\,, 
	\end{aligned}
	\end{cases}
\end{equation}
where -(+) are for vector (axial-vector) current.

\begin{figure}[t]
	\centering
	\includegraphics[width=\columnwidth]{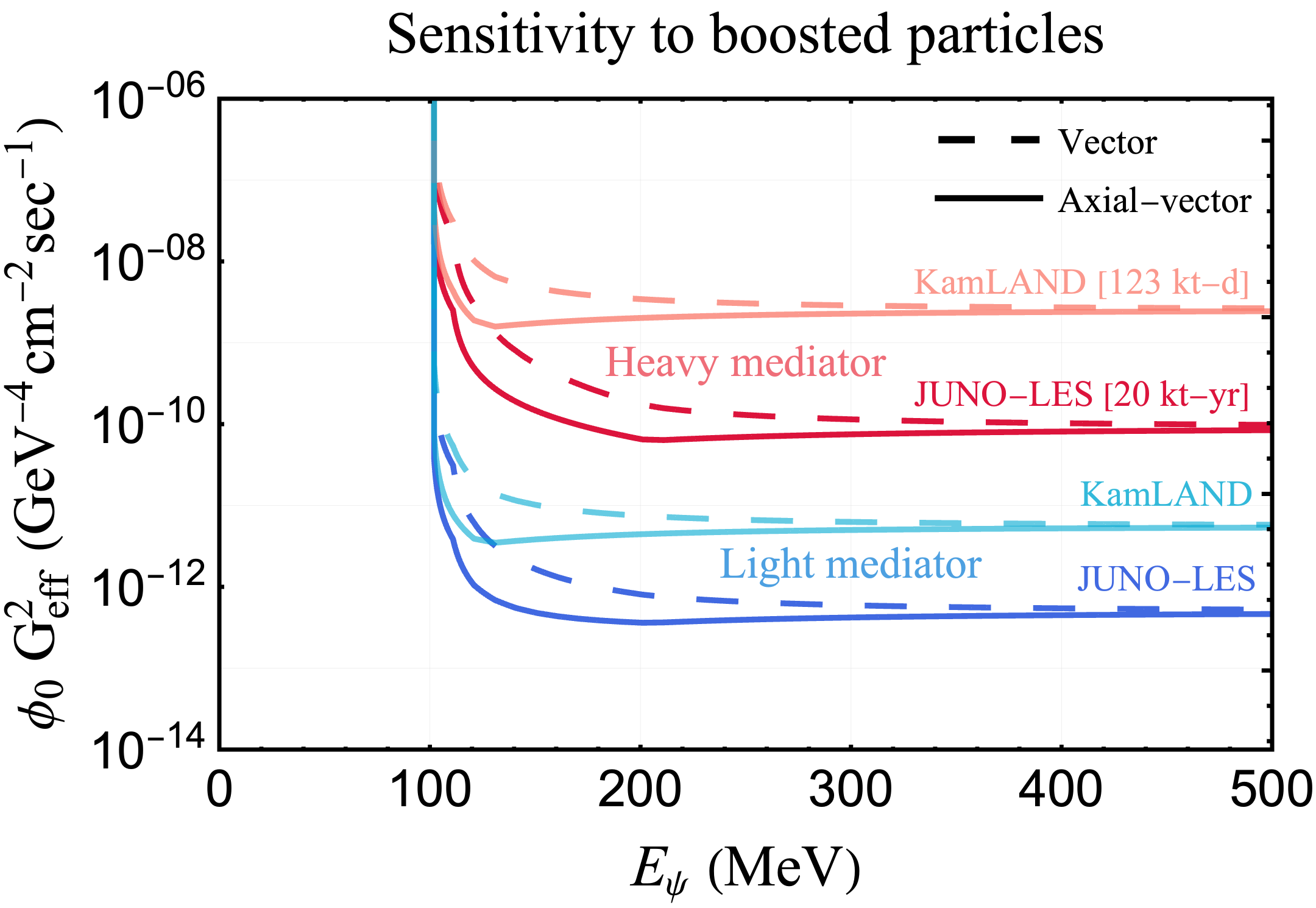}
	\caption{\label{fig:modind} The 90\% C.L. discovery sensitivity of JUNO-LES with 20 kton-yr 
	exposure to the parameters of 
	the monochromatic 
	boosted particle flux is shown for heavy (light) mediator case by red (blue) curves.  The solid 
	(dashed) curve represents the axial-vector (vector) current mediation. The 90\% C.L. exclusion 
	limits obtained from KamLAND with 123 kton-day exposure\,\cite{KamLAND:2011fld} is shown with 
	lighter shades.}
\end{figure}

We can now compute the quenched proton spectrum from $\psi p$ ES. The event rate depends on 
$E_\psi$ and the product $\phi_0 \times G_{\rm eff}^2$. To obtain the 90\% C.L. sensitivity of JUNO to 
the flux of these boosted particles, we consider the events from $\psi p$ ES as signal and the total 
Standard Model {\sc LES} events as the background.  As the spectrum from both interactions is 
predictable, a bin-by-bin comparison will lead to better sensitivity, but is not required for our 
simple analysis. The boosted particle parameter space that can be probed by JUNO is shown in 
Fig.\,\ref{fig:modind}. We find that the event rate is higher for light mediator scenario 
and axial-vector current interaction, as expected.

The volume of KamLAND is 0.697 kton. Using a fiducial exposure of 123 
kton-day, KamLAND has reported one event with $E_{\rm vis} \in$ (13.5, 20) MeV which is consistent 
with their estimate for background\,\cite{KamLAND:2011fld}. The non-observation of excess events in 
this 
bin allows us to put 90\% C.L. exclusion limits on the flux of boosted particles. These limits for the 
various cases are shown in Fig.\,\ref{fig:modind}. The projected sensitivity of JUNO is $\sim 100$ times 
more than KamLAND due to larger exposure and a wider $E_{\rm vis}$ range of the {\sc LES} sample. 

\subsection{Dark matter annihilating to sterile neutrinos}

One of the simplest extensions to the Standard Model is a neutral fermion called 
sterile neutrino ($\nu_s$) which can also act as a portal to dark matter. In these models, the 
annihilation of dark matter is dominated by $\chi \chi \rightarrow \nu_s \nu_s$ which determines the 
relic density\,\cite{McKeen:2018pbb, Kim:2020ipj, Kelly:2019wow}. If the mixing angle between sterile 
and active neutrino is large ($\sim0.1$), the flux of sterile neutrinos will be accompanied by a flux of 
active neutrinos, albeit smaller, which can be detected. However, if 
the mixing angle is small ($<10^{-3}$), the flux of active neutrinos will be too small to be detected and 
one must rely on the detection of $\nu_s$. One of the possibilities is to detect $\nu_s p$ elastic 
scattering 
($\nu_s + p \rightarrow \nu_s + p$) in the {\sc LES} sample at JUNO.

The flux of sterile neutrinos from $s$-channel annihilation of galactic dark matter is given by
\begin{equation}\label{galflux}
	\frac{d\phi}{dE} = \frac{1}{4\pi} \frac{\langle \sigma v \rangle}{M_\chi^2} 
	 \,\delta(E - M_\chi)\,\mathcal{J}\,,
\end{equation}
where $\mathcal{J}\,= \,2.3 \times 10^{23}\,\rm GeV^2\,cm^{-5}$ is the all-sky 
\mbox{$J$-factor}\,\cite{Arguelles:2019ouk}, and $\langle \sigma v \rangle$ is the thermal averaged cross 
section adapted from Ref.\,\cite{Steigman:2012nb}. We have assumed the dark matter to be a 
Majorana fermion. In 
case it were a Dirac fermion, the flux would smaller by a factor of two. To obtain conservative 
estimates, we 
ignore the contribution to flux arising from the extra-galactic component. We also assume that $\nu_s 
p$ ES is mediated by a light vector boson, and take the coupling $g_{\rm L}^{\,} = 0.1$ for illustration. 
 \begin{figure}[h!]
	\centering
	\includegraphics[width=\columnwidth]{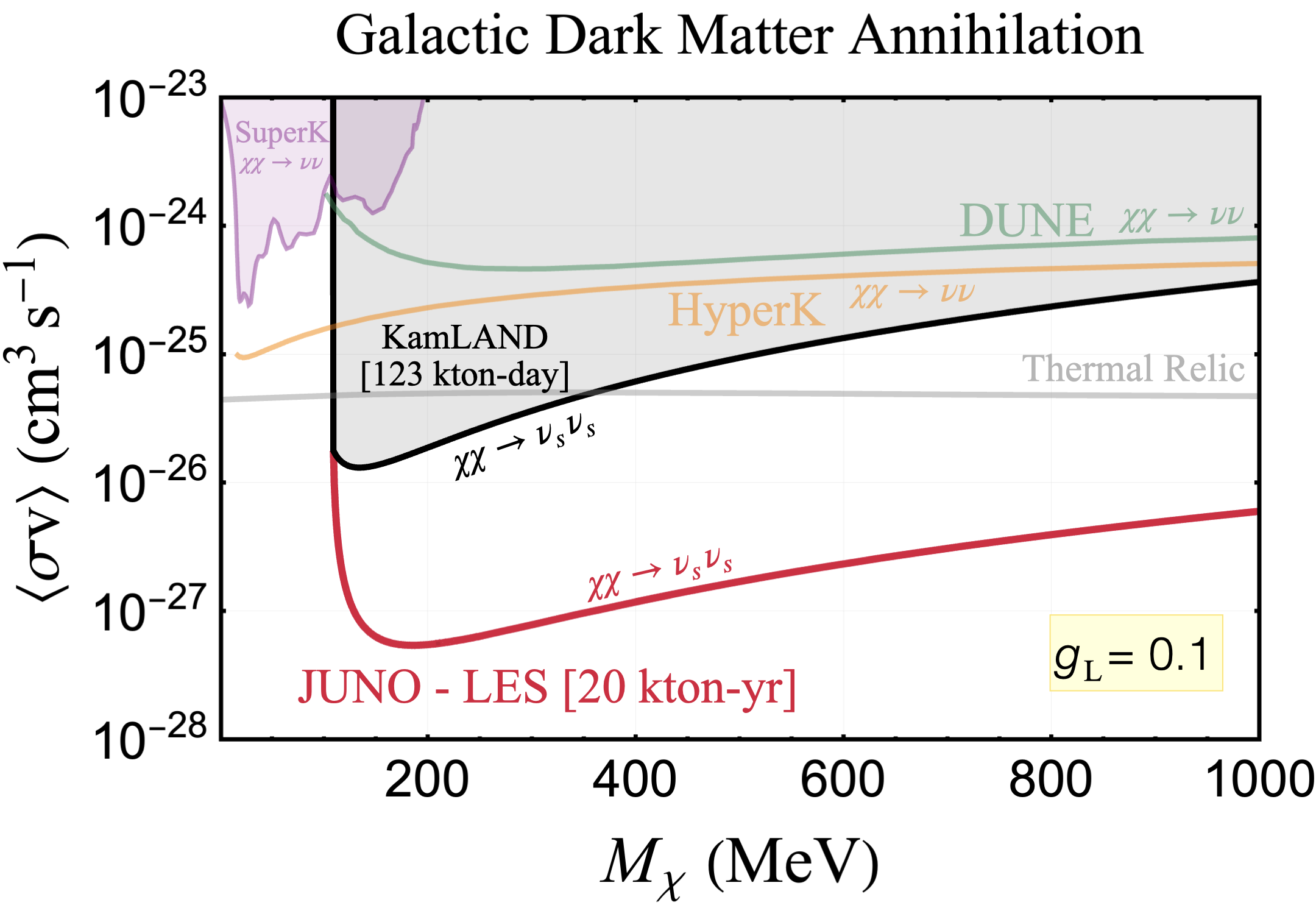}
	\caption{\label{fig:GalDM} The 90\% C.L. exclusions limit from KamLAND\,\cite{KamLAND:2011fld}, 
	and discovery sensitivity 
	of JUNO to the parameters of dark matter annihilation to sterile neutrino, assuming light mediator 
	and $g_{\rm L}^{\,} = 0.1$. The parameter space above the red curve would result in detectable 
	events in the 
	{\sc 
	JUNO-LES} sample with an exposure of 20 kton-yr. The gray curve shows the thermal rate required 
	for obtaining the observed dark 
	matter abundance. For comparison, we show the exclusion limit from Super-Kamiokande obtained 
	from dark matter annihilation to neutrinos \cite{Arguelles:2019ouk, Palomares-Ruiz:2007trf}. We 
	also show the projected sensitivity from DUNE\,\cite{Arguelles:2019ouk} and Hyper-Kamiokande 
	\cite{Bell:2020rkw} to the active neutrino channel. These are relevant for the case of large mixing 
	angle between active and sterile neutrino. }
\end{figure}

The parameter space excluded by KamLAND data, and the 90\% C.L. discovery sensitivity of the {\sc 
LES} 
sample at JUNO with 20 kton-yr exposure are shown in Fig.\,\ref{fig:GalDM}. We also 
show the 
thermal averaged cross section for obtaining the correct relic abundance, and find that JUNO can 
probe this model for dark matter mass in the range 100 MeV to a few GeV. We also show the 
exclusion limits from Super-Kamiokande and the projected sensitivity of DUNE and Hyper-Kamiokande 
for 
dark matter annihilation to active neutrinos, relevant when the mixing angle between active and 
sterile neutrino is large. Note that the above limits were calculated assuming a light mediator. For a 
given value of $M_\chi$, the limits 
for the heavy mediator case can be obtained using Eq.\,\eqref{eq:geff}, which gives the substitution rule
\begin{equation}
	\frac{g_{\rm L}^4}{M_p^4}
	\langle \sigma v \rangle_{\rm L} \leftrightarrow 	\frac{g_{\rm H}^4}{M_{Z^\prime}^4} \langle \sigma 
	v \rangle_{\rm H}, 
\end{equation}
where the subscript L(H) denote the parameters in the light (heavy) mediator scenario. 

\subsection{Boosted Dark Matter}
The elastic scattering between dark matter and proton ($\chi + p \rightarrow \chi +p$) in scintillator 
detector like JUNO  
will lead to a singles event from the scintillation signal of the recoiled proton. In general, the 
momentum transfer to protons from the scattering with cold and non-relativistic dark matter is 
small, quenched, and cannot be detected. However, the interaction of cosmic rays on galactic dark 
matter can \emph{boost} 
these particles to higher velocities, which allows for larger momentum transfers in a detector
\cite{Bringmann:2018cvk,Cappiello:2018hsu}. In Ref.\,\cite{Cappiello:2019qsw}, constraints on the
interaction cross section 
$\sigma_{\chi p}$ and the mass of dark matter $M_\chi$ were obtained from neutrino 
experiments. The sensitivity of JUNO was projected by appropriate scaling of KamLand data. 
The {\sc LES} spectrum computed in this paper will act as a background in the search for such 
boosted dark matter. We recalculate the projected sensitivity of JUNO to $\sigma_{\chi p}$ in the light 
of the LES background.

\begin{figure}[t!]
	\centering
	\includegraphics[width=\columnwidth]{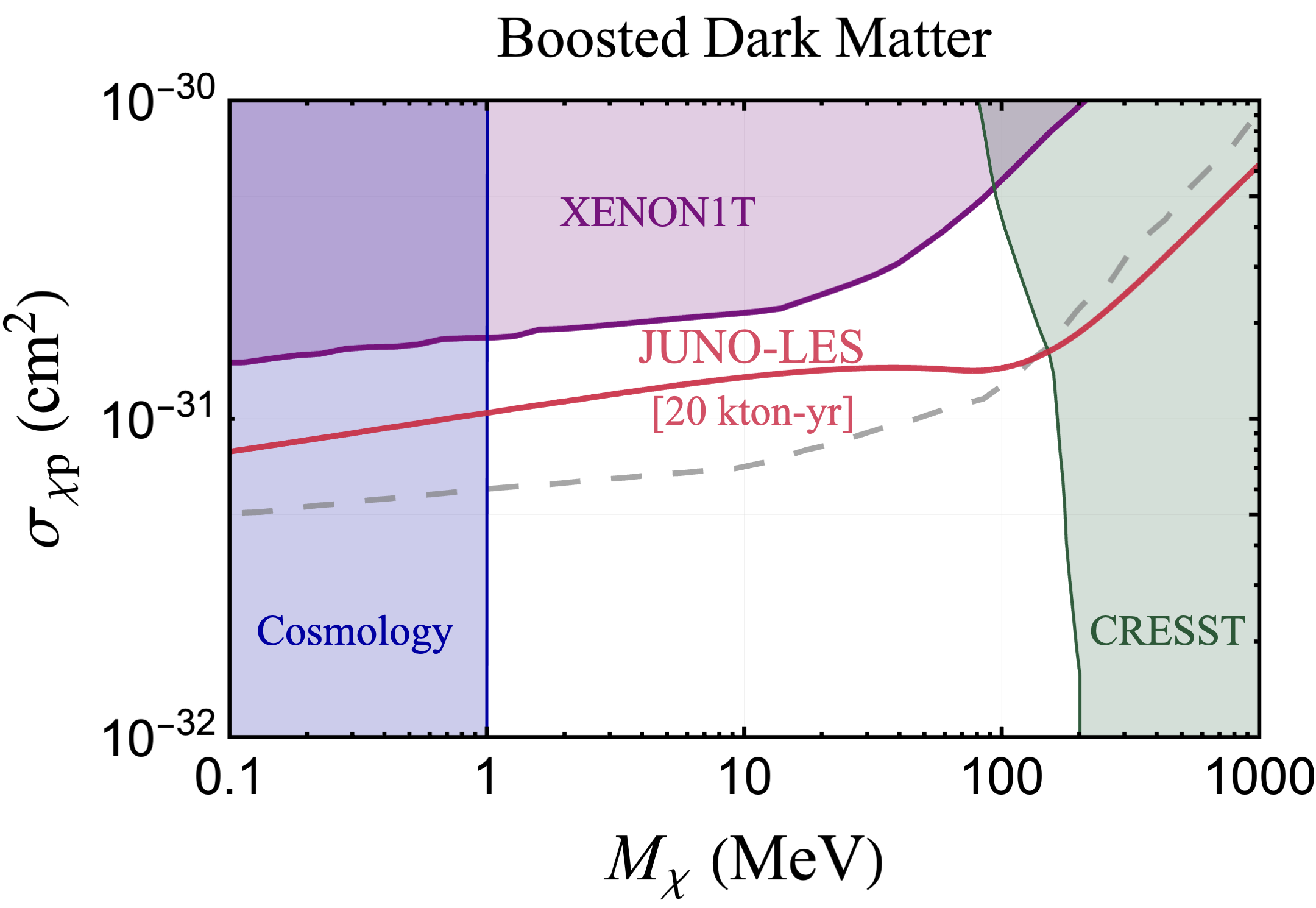}
	\caption{\label{fig:BDM}\,The 90\% C.L discovery sensitivity with the {\sc JUNO-LES} 20 kton-yr 
	sample to the parameters of boosted dark matter is shown in red. The gray 
	dashed line is the corresponding estimate from Ref.\,\cite{Cappiello:2019qsw}. 
	The exclusion limits from XENON1T and other direct detection experiments are adapted from 
	\cite{Bringmann:2018cvk,Cappiello:2018hsu}. The limit from cosmology is taken 
	from\,\cite{Krnjaic:2019dzc}.}
\end{figure}

As the main objective of this paper is to highlight the importance of measurement of the {\sc LES}
sample, we do not compute the flux of boosted dark matter in detail. In Ref.\,\cite{Cappiello:2019qsw}, 
the differential flux is provided for a few benchmark values of dark matter mass. By fitting to a broken 
power law, we find that, for dark matter masses below a few hundred MeV, the flux can be 
approximated by a simplified 
expression
\begin{equation}\label{eq:bdmflux}
	T_\chi \frac{d\phi_\chi}{d T_\chi} \approx \frac{2.5\times10^{-5}}{\rm cm^2\,sec} 
	\left[\frac{\sigma_{\chi 
			p}}{10^{-30}\,\rm cm^2}\right] \left[\frac{\rm 
		MeV}{M_\chi}\right] \left[\frac{T_\chi}{M_\chi} \right]^{\gamma}\,, 
\end{equation}
where
\begin{equation}
	\gamma = 
	\begin{cases}
		\begin{aligned}
			3/4 &\quad  \left[ \, T_\chi<M_\chi \ \, \right]\\
			-3/4 &\quad  \left[ \, T_\chi \geq M_\chi \ \, \right].
		\end{aligned}
	\end{cases}
\end{equation}
The differential cross section for $\chi p$ ES is taken to be $d \sigma_{\chi 
p}/d T_p = \sigma_{\chi p}/T_p^{\rm max}$\,\cite{Cappiello:2019qsw}. We compute the total number of 
events with $E_{\rm vis}\in$(15, 100) MeV, and obtain the 90\% C.L. discovery sensitivity of JUNO by 
comparing with the total events in the {\sc LES} sample, which we consider as background. The results 
are shown in Fig.\,\ref{fig:BDM} along with other relevant 
limits. Our projected sensitivity is consistent with \cite{Cappiello:2019qsw} at higher DM 
masses, however it indicates a degradation in sensitivity for $M_\chi < 100\,\rm MeV$. 

\section{Summary and Outlook}\label{conc}

The neutral-current interactions of atmospheric neutrinos in a large volume liquid scintillator detector, 
such as JUNO, is mainly through elastic scattering (ES) on protons and quasi-elastic-like (QEL) 
scattering on carbon nuclei. The recoiled protons are detected through their scintillation signal. Such 
prompt-only events are also called as ``singles". In this paper, we predict the visible energy 
distribution of singles at JUNO, due to atmospheric neutrino interactions through the $\nu p$ ES and 
$\nu \rm 
C$\,QEL channels. 

We  determine the background due to cosmogenic isotope decay, which would 
dominate for $E_{\rm vis} \leq 16.5$ MeV. Using veto on singles in the vicinity of a muon track, we 
show that the threshold may be reduced to $E_{\rm vis} \sim 15$ MeV, above which the atmospheric 
neutrino signal dominates. Based on our estimates, we propose that 
JUNO can maintain a Large Energy Singles ({\sc LES}) database (i.e., $E_{\rm vis} \geq 15$\,MeV and 
no 
delayed neutron capture signal) wherein the neutral-current interactions of atmospheric neutrinos can 
be detected. The main results of this paper are shown in Fig.\,\ref{fig:result}.

The first goal with the LES events would be to establish their existence, and therefore ensure the 
detection of neutral-current interactions of low-energy atmospheric neutrinos. Assuming only 
statistical errors and no other 
background, we expect JUNO would discover these events with the exposure of a few tenths kton-yr. 
The next step would be a confirmed detection of $\nu p$ ES events, which is a robust prediction of 
Standard 
Model with small uncertainties. We estimate that JUNO can find evidence of $\nu p$ ES by rejecting 
the QEL-only hypothesis at $3\, \sigma$ ($5\, \sigma$) with 12 (34)\,kton-yr exposure. 

The {\sc LES} database can also probe new physics scenarios, which can give rise to singles in the
detector. The {\sc LES} sample is particularly advantageous if 
the new physics model does not admit charged-current-like interactions, for example, in the case of 
boosted dark sector particles. We have estimated the discovery sensitivity of the {\sc LES} sample for 
such scenarios. We also estimate the discovery sensitivity for two well-motivated 
new physics scenarios -- dark matter annihilation to sterile neutrinos, and boosted dark 
matter. In principle, the {\sc LES} sample would also be sensitive to neutral-current non-standard 
interactions that 
modify the predictions of $\nu p$ ES and $\nu \rm C$ QEL channels. However, we have not considered 
this possibility in this work. 

The estimates obtained in the paper are promising. Future work with detailed studies of neutrino 
interactions in JUNO detector will shed more light on backgrounds, and 
aid in developing mitigation techniques. We look forward to a detailed study of the muon spallation at 
JUNO, and veto analysis including pulse shape discrimination. This will allow for a lower threshold, 
and therefore, enhance the prospects for the detection of low-energy atmospheric neutrinos and 
possible new physics signals. \\

\section*{Acknowledgements}
B.C. would like to acknowledge private correspondence with Jie Cheng regarding the results of 
their paper. A.D. would like to thank S. Agarwalla for useful discussions. The work of B.D. and A.D. are 
supported by the Department \,\,of Atomic Energy (Govt.\,\,of India) 
research project under Project Identification No. RTI 4002. B.D. is also supported by a Swarnajayanti 
Fellowship of the Department of Science and Technology (Govt. of India), and by the 
Max-Planck-Gesellschaft through a Max Planck Partner Group.

\onecolumngrid
\newpage
\section*{Supplementary Plots: Vetoing Cosmogenic Backgrounds}
\begin{figure}[h!]
	\centering
	\includegraphics[width=0.49\columnwidth]{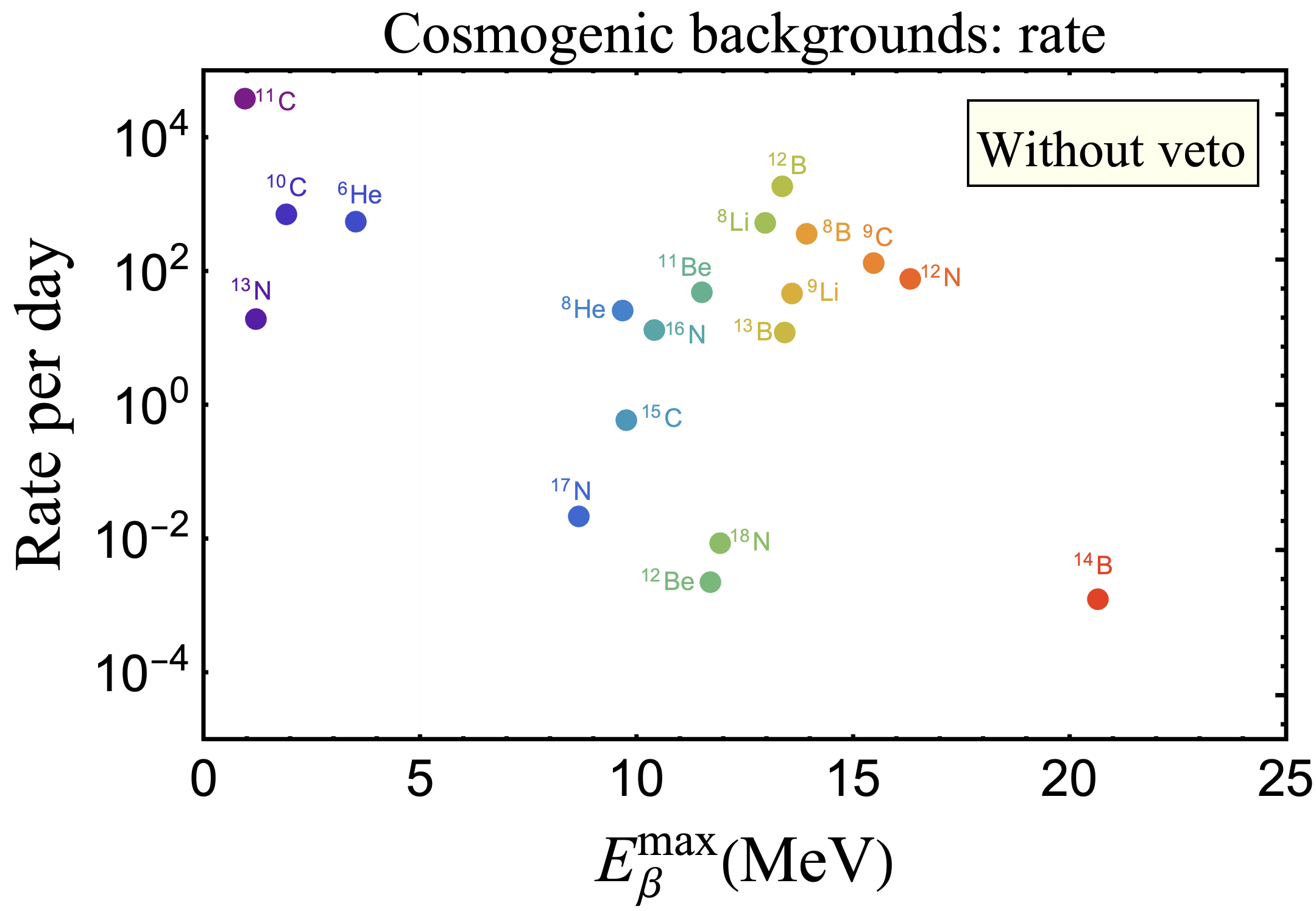}\vspace{0.5cm}
	\includegraphics[width=0.49\columnwidth]{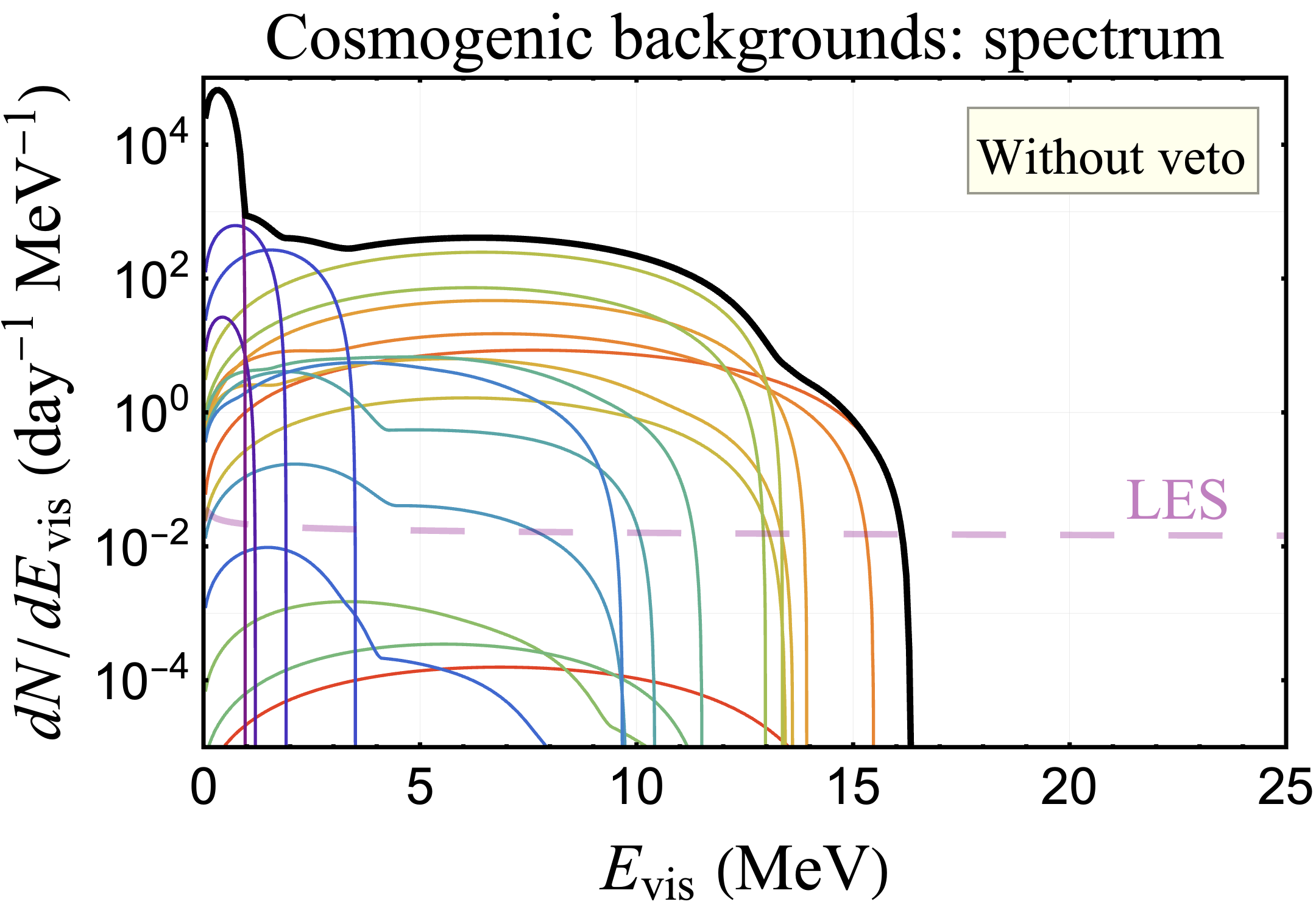}
	\includegraphics[width=0.49\columnwidth]{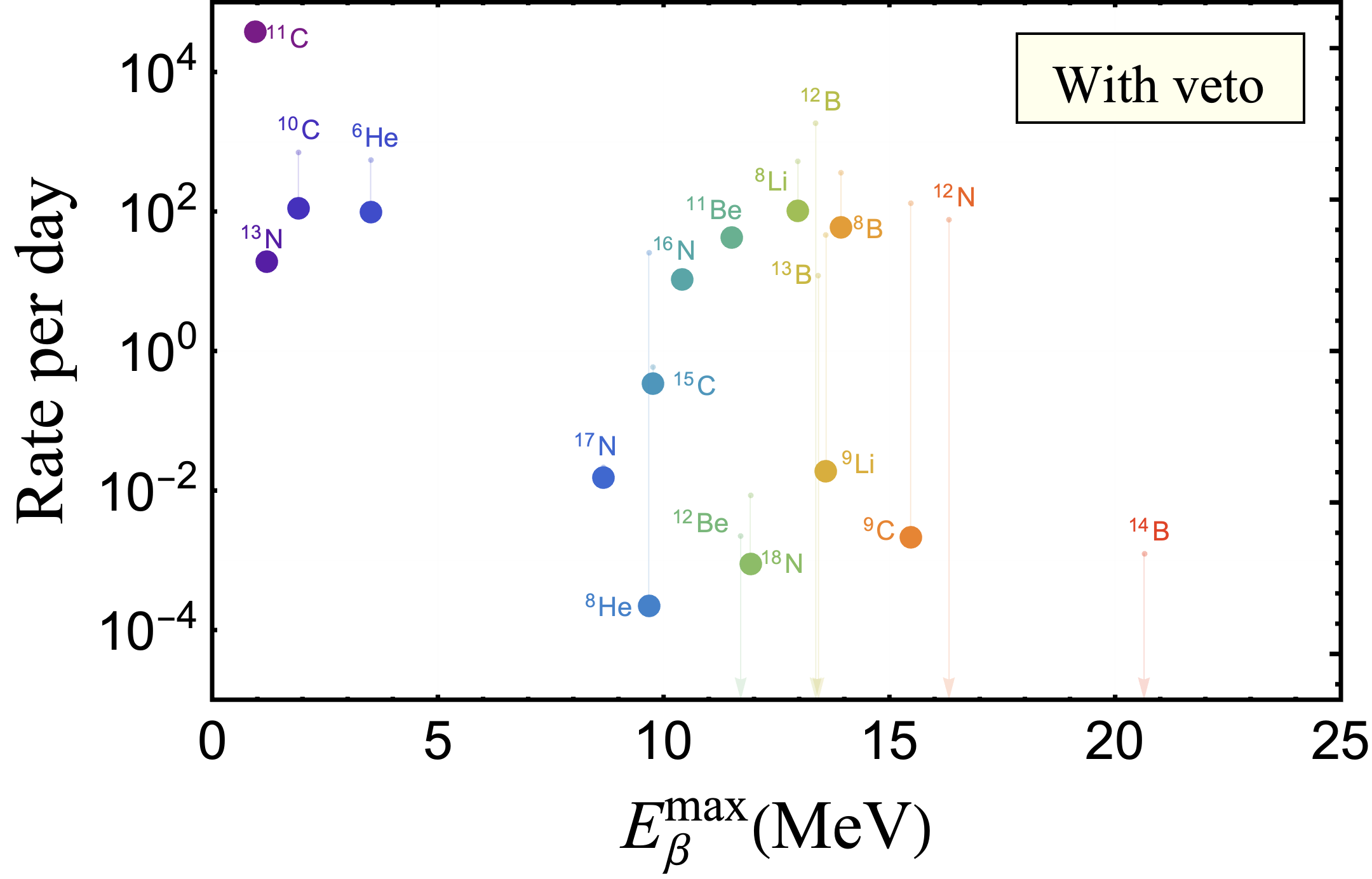}\vspace{0.5cm}
	\includegraphics[width=0.49\columnwidth]{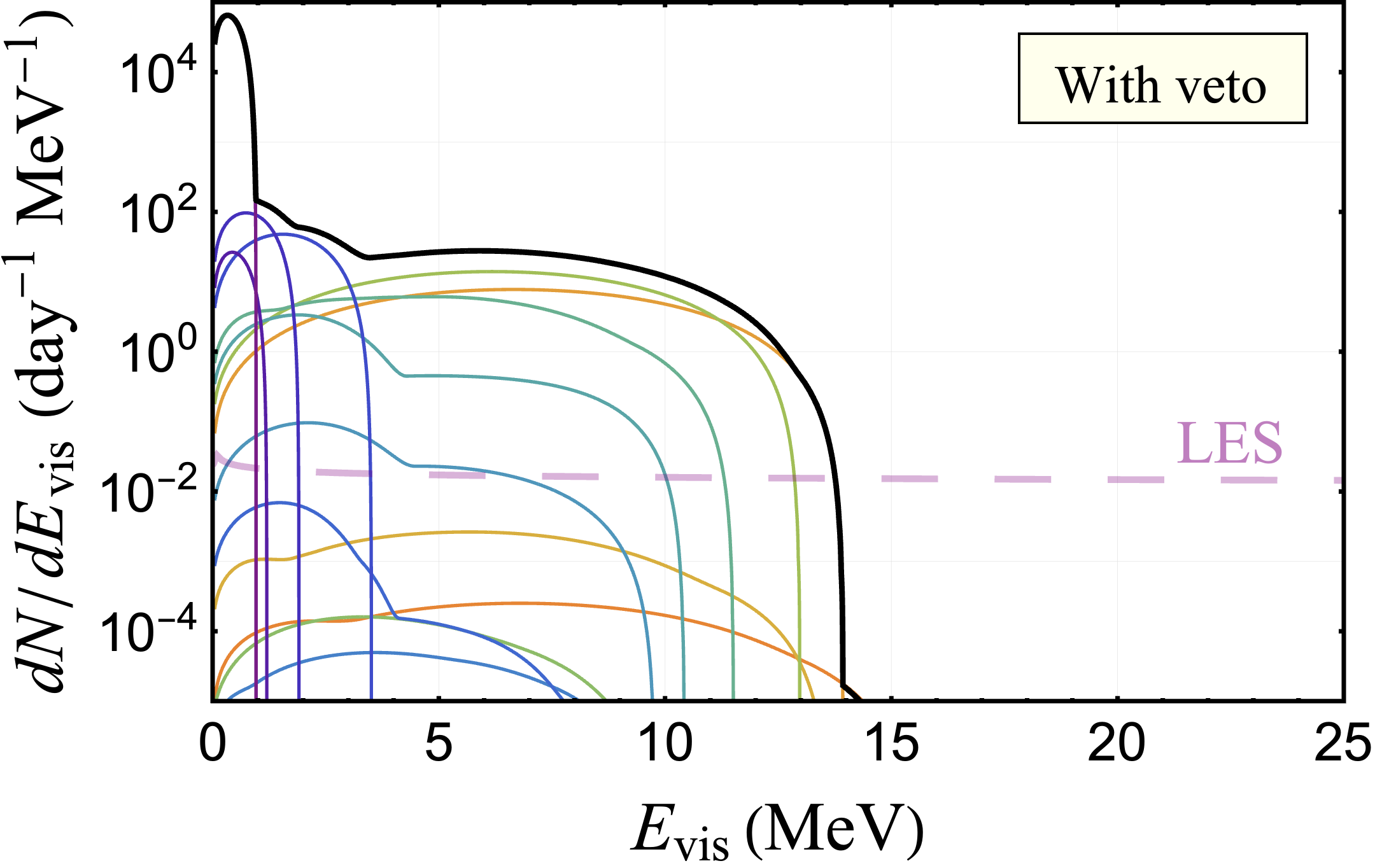}
	\caption{\label{fig:wo} \emph{Left:} The event rate per day for the cosmogenic isotopes is shown 
	against the end point of their beta decay spectrum without (Top) and with (Bottom) $\Delta 
	t_{\rm veto}$ = 2\,s. \emph{Right:} The singles spectrum from decays of cosmogenic isotopes
	without (Top) and with (Bottom) $\Delta t_{\rm veto}$ = 2\,s. The color convention is such that 
	the isotopes with lower (higher) $E_\beta^{\rm max}$ are shown with blue (red) shades. The 
	black curve represents the total cosmogenic background. We have also shown the LES 
	spectrum with dashed light-purple curve for comparison.}
\end{figure}

\newpage
\twocolumngrid

\bibliography{nup_references.bib}

\end{document}